\documentclass[a4paper,11pt]{article}
\usepackage[skins]{tcolorbox}
\usepackage{mathtools,slashed}
\usepackage[T1]{fontenc}
\usepackage{slashed,verbatim,subfigure}
\usepackage[numbers,sort&compress]{natbib}
\usepackage{amsmath}
\usepackage{xcolor}
\usepackage{braket}
\usepackage{mathrsfs}
\usepackage{titlesec}
\usepackage{eucal}
\usepackage{amsbsy}
\usepackage[utf8]{inputenc}
\usepackage{amssymb}
\usepackage{mathtools}
\usepackage{appendix}
\usepackage{graphicx}
\usepackage[final]{pdfpages}
\usepackage{dcolumn}
\usepackage{bm}
\usepackage[colorlinks=true,linktocpage=true,
linkcolor=blue,citecolor=blue]{hyperref}
\usepackage[a4paper]{geometry}
\catcode`@11
\def\seceqaa{\@addtoreset{equation}{section}
	\def\theequation{A\arabic{equation}}}
\def\seceqbb{\@addtoreset{equation}{section}
	\def\theequation{B\arabic{equation}}}
\def\seceqcc{\@addtoreset{equation}{section}
	\def\theequation{C\arabic{equation}}}
\def\seceqdd{\@addtoreset{equation}{section}
	\def\theequation{D\arabic{equation}}}
\def\seceqee{\@addtoreset{equation}{section}
	\def\theequation{E\arabic{equation}}}
\catcode`@11
\newcommand{\be}{\begin{eqnarray}}
	\newcommand{\ee}{\end{eqnarray}}
\usepackage{authblk}
\begin{document}
	\large
	\title{Dirac Brackets $\leftrightarrow$ Lindblad Equation: A Correspondence}
	\author{Aleek Maity\footnote{email- aleek@cmi.ac.in}\vspace{0.1in}~~and V V Sreedhar\footnote{email- sreedhar@cmi.ac.in}}
	\affil{Chennai Mathematical Institute, Plot H1, SIPCOT IT Park, Siruseri\\
		Chennai, India 603103}
	
	\date{}
	
	\maketitle
	\titleformat{\section}
	{\normalfont\Large\bfseries}{\thesection}{1em}{}[{\titlerule[0.8pt]}]
	
	\begin{abstract}
		The time evolution of an open quantum system is governed by the Gorini-Kossakowski-Sudarshan-Lindlad equation for the reduced density operator of the system. This operator is obtained from the full density operator of the composite system involving the system itself, the bath, and the interactions 
		between them, by performing a partial trace over the bath degrees of freedom. The entanglement between the system and the bath leads to a generalized Liouville
		evolution that involves, amongst other things,  dissipation and decoherence of 
		the system. 
		
		In a similar fashion, the time evolution of a physical observable in a classically constrained dynamical system is governed by a generalization of the Liouville equation, in which the usual Poisson bracket is replaced by the so-called Dirac bracket.   The generalization takes into account the reduction in the phase space of the system because of constraints which arise either because they are introduced by hand, or because of singular Lagrangians.
		
		We derive an intriguing, but precise classical-quantum correspondence between the aforementioned situations which connects the Lindblad
		operators to the constraints. The correspondence is illustrated in the ubiquitous example of coupled simple harmonic oscillators studied earlier in various contexts like quantum optics, dissipative quantum systems,  Brownian motion, and the area law for the entropy of black holes.
		
	\end{abstract}
	
	

	\newpage
	\tableofcontents
	
	\section{Introduction}
	Physical systems at the atomic, nuclear, and sub-nuclear levels are governed by quantum mechanics. Since quantum mechanics is an inherently abstract and non-deterministic theory, the correspondence between quantum mechanics and the physically more relatable classical mechanics has intrigued people for several decades. The Correspondence Principle of
	Bohr \cite{bohr}, the Ehrenfest Theorem \cite{ehrenfest}, the Coherent States 
	of Schrodinger \cite{coherent}, and the JWKB approximation \cite{jwkb1} \cite{jwkb2} \cite{jwkb3} \cite{jwkb4} are amongst the earlier attempts to bridge the gap between classical and quantum theories. A little later, Wigner \cite{wigner} invented a quasiprobability distribution using which the expectation values of operators in quantum mechanics could be computed by simply performing averages of the corresponding observables in the classical 
	phase space. 
	
	Another well-known connection between classical and quantum mechanics is based on the prescription of promoting the classical observables to quantum operators \cite{weinberg}. The Poisson brackets of  observables in Hamiltonian mechanics are then replaced by commutation relations between the corresponding operators in quantum mechanics multiplied by $i\hbar$. One may also view quantum mechanics as a purely classical, but suitably geometrized theory \cite{kibble} \cite{heslot}.
	
	A particularly important set of problems is encountered when the dynamics in classical mechanics is confined to a subspace of the 
	phase space. This could be because of constraints introduced by hand, or because the canonical coordinates used to describe the system have some redundant degrees of freedom. Such systems are called constrained systems. Dirac \cite {dirac}\cite{Weinberg}\cite{Hanson}\cite{Mukunda}\cite{KS}\cite{Henneaux} showed how the usual Hamiltonian methods could be generalized to such situations. A central result of this analysis uses the Dirac bracket instead of the Poisson bracket to 
	capture the effect of the constraints in classical mechanics. Canonical quantization consists of replacing the Dirac 
	bracket with a commutator multiplied by $i\hbar$, modulo operator ordering problems.   
	
	Similarly, another important set of problems is encountered when one considers open quantum systems\cite {breuer}\cite{Rivas}\cite{lidar}. In this case, one starts with the total Hilbert Space of the system and the environment. The state of the quantum system is obtained by performing a partial trace over the environment.
	
	Both these problems involve a reduction in the degrees of freedom defining the underlying spaces, one the phase space of classical mechanics, and the other, the Hilbert space of quantum mechanics. In the spirit of the aforementioned correspondence between classical and quantum mechanics, it is therefore natural to examine the correspondence between constrained dynamical systems and open quantum systems.     
	
	In this paper, we do a detailed study of this correspondence. We construct a map between the so-called Lindblad operators \cite{L original}\cite{GKS original}\cite{Manzano} appearing in the Lindblad equation governing open quantum systems on the one hand, and the constraints, appearing through the Dirac bracket, in the dynamical equation of motion governing the corresponding classical system on the other hand. 
	
	The rest of the paper is organized as follows: In Section \ref{S2}, we give a lightning 
	review of constrained dynamical systems emphasizing mainly the definition of Dirac Bracket and the time-evolution of phase space observables in the presence of constraints. The time-evolution equation in open quantum systems is mentioned next, both in terms of the fixed operator basis, and in terms of the Lindblad operators. Finally, we present our main result 
	establishing the advertised correspondence.
	
	In Section \ref{S3}, we illustrate our correspondence in a system of two coupled simple harmonic oscillators. This ubiquitous example appears in various physical contexts in open quantum systems like quantum optics\cite{Carmichael}, Brownian motion \cite{Langevin} and in the context of black hole entropy, where it was earlier studied by Bombelli, Koul, Lee, and Sorkin \cite {bombelli}, and independently by Srednicki \cite{srednicki}. 
	
	In Section \ref{S4} we summarise the paper's results and present an outlook.
	
	A detailed explanation of the Dirac's method for constrained systems and the derivation of the coarse-grained Lindblad equation with the dissipation parameters can be found in Appendix [\ref{DCSA}] and Appendix [\ref{LindbladB}] respectively.

	We use the following notations:
	\text{Poission Bracket} : $\bigr\{,\bigr\}_P$,
	\text{Dirac Bracket}: $\bigr\{,\bigr\}^{\star}$,
	\text{Commutator}: $\bigr[,\bigr]$,
	\text{Anti-commutator}  : $\bigr\{,\bigr\}$. 
	Einstein's summation convention is used, unless explicitly mentioned otherwise. We use `hat' to distinguish quantum observables from their classical counterparts.

	\section{The Dirac Bracket $\leftrightarrow$ Lindblad Equation Correspondence} \label{S2}
	This section consists of three parts. In the first part, we will present a concise review of constrained classical dynamical systems \cite {dirac}\cite{Weinberg}\cite{Hanson} \cite{Mukunda}\cite{KS}\cite{Henneaux}. In the second part, we will briefly describe the master equation governing open quantum systems, namely the Lindblad, Gorini, Kossakowski, and Sudarshan equation, commonly referred to as the Lindblad equation \cite{L original}\cite{GKS original}\cite{Manzano}. In particular we identify the dissipation parameters and the contribution of the Lamb-shift. In the third part, we establish the main result of our paper: The correspondence between Open Quantum Systems and Constrained Classical Dynamical Systems.


	\subsection{Constrained Classical Dynamical Systems:}
	Classical dynamical systems defined on phase spaces in which not all the phase space coordinates are independent are called constrained systems. The relations between the phase space coordinates are described by constraint equations of the form
	\begin{equation}
		\phi_m (q,p) \approx 0.
	\end{equation} 
	These constraints might naturally arise in defining the canonically conjugate momenta in systems with singular Lagrangians, or can be imposed by hand, and are called primary constraints. As may be expected, the dynamics of such systems is restricted to a subspace of the phase space defined by the constraints. This fact is represented by the symbol `$\approx$' (read weakly equal to) to indicate an equality which holds only on the constrained phase space, but not necessarily on the entire phase space.  
	
	The above complications require a generalization of the usual Hamiltonian approach to dynamical systems, developed by Dirac \cite {dirac}\cite{Weinberg}\cite{Hanson} \cite{Mukunda}\cite{KS}\cite{Henneaux}, which we briefly review in this section. We begin by noting that the canonical Hamiltonian is not unique due to the presence of constraints
	\begin{equation}
		H_c + \lambda_m \phi_m \approx H_c
	\end{equation}
	$\lambda_m$ are Lagrange multipliers that enforce the 
	constraints. The requirement that the constraints be preserved in time may lead to more conditions among the phase space variables and hence more constraints $\chi_n (q,p) \approx 0$, called secondary constraints, or they may produce equations which can be solved for one of more of the Lagrange multipliers $\lambda_m$. 
	
	Once all the primary and secondary constraints are 
	determined, they may be classified into two types: first and second-class. By definition, if a quantity has a vanishing Poisson bracket with all constraints, it is first-class, and second-class otherwise. Physically, first class constraints generate gauge transformations, while second-class
	constraints merely reduce the phase space.   
	
	The gauge invariance in the theory can be fixed by an appropriate choice of gauge conditions which in effect convert the first class constraints into second class constraints. At this stage, all the Lagrange multipliers are determined, and standard Hamiltonian methods can be applied by replacing the 
	usual Poisson bracket between two classical observables $R$ and $S$ with the Dirac Bracket defined by
	\begin{equation}
		\bigr\{R,S \bigr\}^{\star} = \bigr\{R ,S  \bigr\}_P -\bigr\{R,\phi_{\alpha}\bigr\}_P ~ C^{-1}_{\alpha\beta} ~ \bigr\{\phi_{\beta},S \bigr\}_P
	\end{equation}
	The time-evolution of a phase space variable $R(q,p)$ can be given by
	\begin{equation}\label{evolution}
		\dot{R}(t)=\bigr\{R,H_c \bigr\}^{\star} \equiv \bigr\{R ,H_c  \bigr\}_P -\bigr\{R,\phi_{\alpha}\bigr\}_P ~ C^{-1}_{\alpha\beta} ~ \bigr\{\phi_{\beta},H_c \bigr\}_P 
	\end{equation}
	where $\phi_\alpha$, $\phi_\beta$ are the second class constraints and $C_{\alpha\beta} \equiv \bigr\{\phi_\alpha ,\phi_\beta \bigr\}_P$ is the matrix of Poisson brackets of second class constraints.
	
	
	\subsection{Lindblad Equation:}
	An open quantum system is described by a Hamiltonian which consists of three parts: A system Hamiltonian($H_S$), a bath Hamiltonian($H_B$), and an interacting part($H_I$). The master equation governing the time evolution of the reduced density matrix (obtained by taking a partial trace over the bath degrees of freedom of the total density matrix) can be obtained from the unitary evolution of the total density matrix taken to be a separable product state of the
	form $\hat{\rho}_{T} = \hat{\rho}_{S}\otimes\hat{\rho}_{B}$. 
	
	The state of the reduced system is given by the Kraus Operator Sum Representation (OSR)\cite{lidar} \cite{Manzano}\cite{Nielsen}
	\begin{equation}\label{OSR}
		\hat{\rho}_{s}(t)  =\sum_{lm}\hat{K}_{lm}(t)\hat{\rho}_{s}(0)\hat{K}^{\dag}_{lm}(t)  
	\end{equation}
	where $\hat{K}_{lm}(t)$ are the Kraus operators that act on the Hilbert space of the reduced system $\mathcal{H_S}$ [eq.(\ref{kraus})]. $l,m$ take values $0,1,2,...,d_B-1$, $d_B$ being the dimension of the bath's Hilbert space $\mathcal{H_B}$.
	
	The Kraus operators may be expanded in the (time-independent) fixed operator basis $\bigr\{\hat{S}_\alpha \}^{d^2_{s}-1}_{\alpha =0}$ of the system Hilbert space $\mathcal{H_S}$,
	$\Bigl\{\hat{S}_\alpha \Bigl\}^{d^{2}_{S}-1}_{\alpha =0}$ with $\hat{S}_0 = \hat{I}$,
	\begin{equation} \label{k-sys}
		\hat{K}_{i}(t)=\sum_{\alpha =0}^{d^{2}_{S}-1} b_{i \alpha}(t) \hat{S}_\alpha
	\end{equation}
	where, $b_{i\alpha}$ are the time-dependent elements of a rectangular $d^{2}_{S}\times d^{2}_{B}$ dimensional matrix. Substituting eq.(\ref{k-sys}) in eq.(\ref{OSR}) 
	\begin{equation}\label{WWE}
		\hat{\rho}_s (t) =\chi_{00}(t) \hat{\rho}_s (0)+\sum_{\alpha \geq 1} \bigr[\chi_{0\alpha}(t)\hat{\rho}_s (0)\hat{S}_{\alpha}^{\dag}+\chi_{\alpha 0}(t)\hat{S}_{\alpha}\hat{\rho}_s (0)\bigr]+\sum_{\alpha,\beta \geq 1} \chi_{\alpha\beta}\hat{S}_{\alpha}\hat{\rho}_s (0)\hat{S}_{\beta}^{\dag}
	\end{equation}
	where
	\begin{equation} \label{chi}
		\chi_{\alpha\beta}(t)=\sum_{i=0}^{d_{B}^{2}-1} b_{i\alpha}(t)b^{\star}_{i\beta}(t)
	\end{equation}
	$\chi(t)$ is a positive semi-definite, hermitian $d^{2}_{s} \times d^{2}_{s}$ matrix
	which is related to dissipation effects. 
	
	After appropriate approximations, the time-evolution equation can be brought to the form
	\begin{equation} \label{CG-SP}
		\frac{d}{dt}\hat{\rho}_{s}(t)=-i\Bigr[\hat{H}_{S}+\hat{H}_{LS},\hat{\rho}_{s}(t)\Bigr]+\sum_{\alpha,\beta \geq 1} \gamma_{\alpha\beta} \biggl(\hat{S}_{\alpha}\hat{\rho}_{s}(t)\hat{S}^{\dag}_{\beta}-\frac{1}{2}\Bigl\{\hat{S}^{\dag}_{\beta}\hat{S}_{\alpha},\hat{\rho}_{s}(t)\Bigl\}\biggl)
	\end{equation}
	The various approximations\cite{lidar}\cite{CG original}\cite{LBW} that go into the derivation of the above equation from the eq.(\ref{WWE}) are explained in the Appendix [\ref{LindbladB}]. The
	integration over the bath degrees of freedom results in two significant contributions to the system dynamics {\it viz.} a contribution to the unitary evolution governed by the the Lamb-shift Hamiltonian($H_{LS}$) and dissipation effects captured by the parameters 
	\begin{equation} \label{d-parameter}
		\gamma_{\alpha\beta}=\langle \dot{\chi}_{\alpha\beta}\rangle=\frac{\chi_{\alpha\beta}(\tau)}{\tau}=\tau \sum_{\alpha^{\prime}\alpha^{\prime\prime} \alpha^{\prime\prime\prime}
			\beta^{\prime}\beta^{\prime\prime}\beta^{\prime\prime\prime}   }\lambda_{\alpha^{\prime\prime} \alpha^{\prime\prime\prime}} 
		\Bigl(\lambda_{\beta^{\prime\prime} \beta^{\prime\prime\prime}}\Bigl)^{\star}\langle \hat{B}_{\alpha^{\prime\prime}}\hat{B}^{\dag}_{\beta^{\prime\prime}}\rangle_{B} \Gamma^{\alpha \alpha^{\prime}}_{\alpha^{\prime\prime} \alpha^{\prime\prime\prime}}(\tau) \Bigl(\Gamma^{\beta \beta^{\prime}}_{\beta^{\prime\prime} \beta^{\prime\prime\prime}}(\tau))\Bigl)^{\star}
	\end{equation}
	where $\lambda$ are the coupling constants between the system and bath, $\hat{B}$ are the bath operators and $\Gamma_{\beta\delta}^{\alpha\gamma}$ are given in terms of integrals over the time-dependent coefficients of the system and bath operators.
	 
	For the sake of completion, we mention that diagonalizing the dissipation parameters allows us bring the eq.(\ref{CG-SP}) to the Lindblad form
	\begin{equation} \label{CG-L}
		\frac{d}{dt}\hat{\rho}_{s}(t)=-i\Bigr[\hat{H}_{S}+\hat{H}_{LS},\hat{\rho}_{s}(t)\Bigr]+\sum_{c \geq 1}\tilde{\gamma}_{c} \biggl(\hat{L}_c \hat{\rho}_{s}(t)\hat{L}^{\dag}_{c}-\frac{1}{2}\Bigl\{\hat{L}^{\dag}_{c}\hat{L}_{c},\hat{\rho}_{s}(t)\Bigl\}\biggl)
	\end{equation}
	where $\tilde{\gamma}$ are the eigenvalues of $\gamma$ and $\hat{L}$ are the Lindblad operators.
	
	
	\subsection{The Correspondence:}
	The evolution equation for the density function in a classical constrained dynamical system is given by [eq.(\ref{evolution})]
	\begin{equation}\label{rho1}
		{d\over{dt}} \rho(t)= \bigr\{ \rho, H_c \bigr\}^{\star}= \bigr\{ \rho,H_c  \bigr\}_P-\bigr\{ \rho,\phi_{a}\bigr\}_P ~D_{ab}~ \bigr\{ \phi_{b},H_c \bigr\}_P 
	\end{equation} 
	where,
	\begin{equation}
		D_{ab} = C^{-1}_{ab},\qquad  C_{ab}=\bigr\{\phi_a,\phi_b \bigr\}_P
	\end{equation}
	and $\phi_a$ are the second-class constraints of the system.
	
	Since all the constraints have been accounted for, $\bigr\{ \phi_{b},H_c \bigr\}_P $ can not produce new constraints. Denoting it by $\chi_b \equiv \dot\phi_b$, it has to be weakly zero, and hence a linear combination of constraints
	\begin{equation}
		{d\over{dt}} \rho(t)= \bigr\{ \rho,H_c  \bigr\}_P - \bigr\{ \rho,\phi_{a}\bigr\}_P D_{ab} \chi_b
	\end{equation}
	Assuming there are no operator ordering problems, we can promote the phase space observables (and hence the constraints) to quantum operators and replace the Poisson Brackets with commutators\footnote{Promoting the Dirac bracket to a commutator enforces the weak equality at the quantum level. Alternatively, one can use the classical 
	Dirac bracket expression and promote the 
	Poisson brackets that appear in its definition to commutators. The two approaches give different results in general. However, restricting to the constrained surface in the classical theory, and acting on the physical states in the quantum theory, the two approaches yield the same result. 
	
	It is worth recalling that in the canonical quantization of gauge theories, there are two ways to translate the weak equality satisfied by the constraints to quantum theory. First using Dirac brackets instead of Poisson brackets and formally promoting classical observables to quantum operators automatically enforces the weak equality. Another way is to introduce a 
	Hilbert Space ${\cal H}^{'}$, with possibly a non-positive-definite metric, with operators corresponding to each unconstrained observable, and standard commutation relations between them derived from Poisson brackets. The constraints act as non-vanishing operators in this Hilbert Space, with vanishing matrix elements between physical states. 
	
	For example, in the Gupta-Bleuler quantization of Maxwell theory, the Lorentz gauge condition is enforced in the quantum theory by the requirement that $\bra{\chi}\partial_\mu A^\mu\ket{\psi} = 0$, for physical states $\ket{\psi}$ and $\ket{\chi}$ \cite{Zuber}.
	
	A stronger statement consists in saying that the physical Hilbert Space ${\cal H}$ is obtained as the subspace of states annihilated by the constraint operators.}
	\begin{equation}\label{rho2}
		{d\over{dt}} \hat{\rho}(t)=i \bigr[ \hat{\rho},\hat{H}_c  \bigr] - \bigr[ \hat{\rho},\hat{\phi}_{a}\bigr] D_{ab} \hat{\chi_b}
	\end{equation}
	Using the following commutator and anticommutator
	\begin{equation}
		\begin{split}
			\Bigl\{ \hat{\phi}_a \hat{\rho}, D_{ab}\hat{\chi}_b \Bigl\} &= \hat{\phi}_a \hat{\rho} D_{ab} \hat{\chi}_b +D_{ab}\hat{\chi}_b \hat{\phi}_a \hat{\rho} \\
			\Bigl[ \hat{\phi}_a , D_{ab}\hat{\chi}_b \Bigl] &= \hat{\phi}_a D_{ab} \hat{\chi}_b -D_{ab} \hat{\chi}_b \hat{\phi}_a
		\end{split}  
	\end{equation}
	Equation (\ref{rho2}) becomes,
	\begin{equation}\label{rho3}
		\dot{\hat{\rho}}(t) = i \bigr[ \hat{\rho},\hat{H}_c  \bigr] - \Bigl\{ D_{ab}\hat{\chi}_b \hat{\phi}_a , \hat{\rho} \Bigl\} + \Bigl\{ \hat{\phi}_a \hat{\rho} ,D_{ab}\hat{\chi}_b \Bigl\} - \hat{\rho} \Bigl[ \hat{\phi}_a ,D_{ab}\hat{\chi}_b  \Bigl]
	\end{equation}
	
	A direct comparison of the above equation with the time-evolution equation [eq.(\ref{CG-SP})] gives,
	\begin{align}\label{correspondence}
		D_{ab} \hat{\chi}_b \hat{\phi}_a =   \frac{1}{2}\gamma_{\alpha\beta}\hat{S}^{\dag}_{\beta}\hat{S}_{\alpha}
	\end{align}

	Equation (\ref{correspondence}) is the central result of this paper.\footnote{The choice of indices in this equation may be confusing, but it is unavoidable. The Latin indices refer to constraints and the Greek indices refer to jump operators, both of which have independent existence. An equation expressing a correspondence between the two {\it a priori} unrelated objects naturally involves both the types of indices.}
	
	The time-evolution equation [eq.(\ref{CG-SP})] is written in the fixed operator basis $\bigl\{ \hat{S}_{a} \bigl\}$.
	To bring it to the Lindblad form[eq.(\ref{CG-L})], we note that $\gamma$ is positive semidefinite and we can diagonalize it as follows:
	\begin{equation}\label{diag_gamma}
		\tilde{\gamma} = u^{\dag} \gamma u
	\end{equation} 
	The RHS of eq.(\ref{correspondence}) becomes,
	\begin{equation}
		\frac{1}{2} \gamma_{ab} \hat{S}^{\dag}_{\;b} \hat{S}_{a} = \frac{1}{2} u_{ac} \tilde{\gamma}_c (u^{\dag})_{cb} \hat{S}^{\dag}_b \hat{S}_a 
		= \frac{1}{2} \hat{L}_c \; \tilde{\gamma}_c \; {\hat{L}^{\dag}_c}
	\end{equation}
	enabling us to identify the Lindblad operators
	\begin{equation}\label{lindbladops}
		\hat{L}_c = \hat{S}_a u_{ac}
	\end{equation}
	Similarly, the real, anti-symmetric matrix $D_{ab}$ on the LHS[eq.(\ref{correspondence})] can be brought to a block diagonal form through an orthogonal transformation
	\begin{equation}
		O^T D O = B
	\end{equation} 
	where $B$ is a block diagonal antisymmetric matrix with entries along the diagonal given 
	by 2x2 antisymmetric matrices as follows:
	\[B = \Biggl\{
	\begin{pmatrix}
		0 & b_1 \\
		-b_1 & 0
	\end{pmatrix}
	,
	\begin{pmatrix}
		0 & b_2 \\
		-b_2 & 0
	\end{pmatrix}
	...
	\begin{pmatrix}
		0 & b_n \\
		-b_n & 0
	\end{pmatrix}
	\Biggl\}
	\]
	Thus the LHS becomes,
	\begin{equation}
		D_{ab} \hat{\chi}_b \hat{\phi}_a = O_{ac} B_{cd} O_{bd} \hat{\chi}_b \hat{\phi}_a =   \hat{\Phi}_{c} B_{cd} \dot{\hat{\Phi}}_d
	\end{equation}
	where we have defined,
	\begin{equation}
		\hat{\Phi}_{c} = O_{ac} \hat{\phi}_{a} \quad {\hbox{and}}\quad \dot{\hat{\Phi}}_d = O_{bd} \hat{\chi}_b , \;\; {\hbox{since}}, ~\hat{\chi}_b = \dot{\hat{\phi}}_b
	\end{equation}
	The correspondence relation in equation (\ref{correspondence}) now takes the form\footnote{It is straightforward to check that the remaining terms on the classical side involving $\hat{\rho}$ automatically match the corresponding terms on the quantum side {\it viz.} $\hat{L}\hat{\rho} \hat{L}^\dagger$, once the stated equality holds and a suitable ordering is prescribed for the operator products when the classical observables are promoted to quantum operators in the standard procedure of canonical quantization.}$~^{,}$ 
	\begin{equation}\label{main c}
		\hat{\Phi}_{c} B_{cd}\; \dot{\hat{\Phi}}_d  =  \frac{1}{2} \hat{L}_c \; \tilde{\gamma}_c \; {\hat{L}^{\dag}_c} 
	\end{equation}
	One can in principle solve the above equation to relate the Lindblad operators to 
	the constraints, and {\it vice versa}\footnote{Note that we can not relate the Lindblad operators to the constraints in 
		a one-to-one
		manner. In general, there are multiple jump operators and constraints, but only one
		condition between them. Nevertheless it may be possible to guess the relation between the jump
		operators and the constraints in specific cases, {\it e.g} coupled
		harmonic oscillators discussed in the next section(\ref{S3}).}. In the next section, we illustrate this in a 
	specific example, {\it viz.}, the case of 
	two coupled simple harmonic oscillators. 
	\section{An example: Two coupled oscillators}\label{S3}
	
	The model we consider was studied earlier by
	Srednicki and Bombelli {\it et al} in the context 
	of the area law for black hole entropy \citep{bombelli}\cite{srednicki}. Their toy model describes two interacting simple harmonic oscillators in one dimension with the
	Hamiltonian
	\begin{equation}
		\begin{split}
			\hat{H} =\frac{1}{2}\hat{p}^2_{1}+\frac{1}{2}k_{1}\hat{x}^{2}_{1}+\frac{1}{2}\hat{p}^2_{2}+\frac{1}{2}k_{2}\hat{x}^{2}_{2}-k^{\prime} \hat{x}_{1} \hat{x}_{2}
		\end{split}
	\end{equation} 
	For simplicity, we choose unit masses for both the oscillators ($m_{1}=m_{2}=1$). The first oscillator, labeled by $1$ is considered as
	the system, and the second oscillator, labeled by $2$ is considered as the bath.\footnote{Usually the bath consists of an infinite number of oscillators. We approximate the bath by a single oscillator to illustrate the main result of this paper. It is not uncommon to represent a bath by a single oscillator: In the Jaynes-Cummings model for atom-radiation interaction, for example, the atom is approximated by a two-level system, and the bath by a single mode of electromagnetic radiation in a 
		cavity \cite{jaynes}.}
	
	The interaction Hamiltonian is
	\begin{equation}\label{int-ham1}
		\hat{H}_{I}=-k^{\prime}\hat{x}_{1}\hat{x}_{2}=\kappa(\hat{a}+\hat{a}^{\dag})\otimes(\hat{b}+\hat{b}^{\dag})
	\end{equation}
	where,
	\begin{align*}
		\kappa &=-\frac{k^{\prime}}{2\sqrt{\omega_{0}\omega_B}}; \qquad k_1 = \omega^{2}_{0}; \qquad k_2 = \omega^{2}_{B}\\
		\hat{x}_1 &=\sqrt{\frac{1}{2\omega_0}} (\hat{a}+\hat{a}^{\dag}), \quad 
		\hat{x}_2 =\sqrt{\frac{1}{2\omega_B}} (\hat{b}+\hat{b}^{\dag})
	\end{align*}
	In the interaction picture eq.(\ref{int-ham1}) becomes,
	\begin{equation}\label{int_ham_2}
		\begin{split}
			\hat{H}_{I}(t)&=\kappa\Bigl(\hat{a}(t)+\hat{a}^{\dag}(t)\Bigl)\otimes\Bigl(\hat{b}(t)+\hat{b}^{\dag}(t)\Bigl)
		\end{split}
	\end{equation}
	This equation is of the general form
	\begin{equation}\label{H_int}
		\hat{H}_{I}(t)=\sum_{i, \;k} \lambda_{ik} \; \hat{S}_{i}(t)\otimes \hat{B}_{k}(t)
	\end{equation}	
	where $i=1,2$ and $k$ refers to the total number of bath oscillators. Since there is only one bath oscillator we will henceforth suppress the index $k$. 
	A direct comparison of the above two equations gives
	\begin{equation}\label{system ops}
		\hat{S}_{i}(t)=\sum_{i^{\prime}} p_{ii^{\prime}}(t)\hat{S}_{i^{\prime}}=\hat{a}_i (t)=\hat{a}_{i}\mathrm{e}^{i\omega_{0i}t} \qquad  
	\end{equation}
	where, $\hat{a}_1 = \hat{a},\; \hat{a}_2 = \hat{a}^{\dag}, \quad
	\omega_{01}=\omega_{0},\quad ~~\omega_{02}=-\omega_{0}$, 
	and
	\begin{equation}
		\hat{B}(t)=  \hat{b}(t)+\hat{b}^{\dag}(t) = q(t) \hat{b} + q^{\star}(t) \hat{b}^{\dag}
	\end{equation} 
	\subsection{Dissipation Parameters and the Lamb Shift}
	Let us split the Hamiltonian [eq.(\ref{int_ham_2})] into two parts, the first one being,  
	$\sum_{i} \kappa \hat{a}_{i}(t) \otimes  \hat{b} \mathrm{e}^{i\omega_B t}$, and the second one $ \sum_{i} \kappa \hat{a}_{i}(t) \otimes  \hat{b}^{\dag} \mathrm{e}^{-i\omega_B t}$. 
	
	Comparing the first part with eq.[(\ref{H_int}),(\ref{system ops})], we get  
	\begin{equation}
		\begin{split}
			\lambda_{i}=\kappa , \quad
			p_{ii^{\prime}}(t)= \delta_{ii^{\prime}} \mathrm{e}^{i\omega_{0i}t} , \quad
			q(t)=  \mathrm{e}^{i\omega_B t} ,	 \quad \hat{S}_{i^{\prime}}=\hat{a}_{i^{\prime}} 
		\end{split}
	\end{equation}
	Let $\alpha \equiv i^{\prime}$ and $\beta \equiv i$ be the system index and the interaction index respectively.
	
	From eq.(\ref{d-parameter}) the dissipation parameter corresponding to the first part is,
	\begin{equation}\label{DP-1}
		\begin{split}
			\frac{\chi^{(1)}_{i^{\prime}j^{\prime}}(\tau)}{\tau}	&=	\tau
			\sum_{i;j} \lambda_{i} {\lambda^{\star}}_{j} \; \langle \hat{b} {\hat{b}^{\dag}}\rangle_{B}\; \Gamma^{i^{\prime}}_{i} (\tau) \Bigl(\Gamma^{j^{\prime}}_{j} (\tau)\Bigl)^{\star} \\
			&= \tau  \lambda_{i^{\prime}} {\lambda^{\star}}_{j^{\prime}} \; \langle \hat{b} {\hat{b}^{\dag}} \rangle_{B}\; \Gamma(\omega_{0i^{\prime}}+\omega_B) \Gamma(-\omega_{0j^{\prime}}-\omega_B)
		\end{split}
	\end{equation}
	where,
	\begin{equation}
		\langle {\hat{b}^{\dag}} \hat{b}\rangle_{B} = \langle \hat{b} \hat{b}^{\dag}\rangle_{B} - 1 = \frac{1}{\mathrm{e}^{\beta \omega_B}-1}
	\end{equation}
	\begin{equation}
		\label{a}
		\begin{split}
			\Gamma^{i^{\prime}}_{i}(\tau) =\frac{1}{\tau}\int_{0}^{\tau}dt \;p_{ii^{\prime}}(t)\; q(t)
			=\delta_{ii^{\prime}}\; \;\Gamma(\omega_{0i}+\omega_B)
		\end{split}
	\end{equation}
	\begin{equation}
		\Gamma(\omega_B)=\frac{1}{\tau}\int_{0}^{\tau}\mathrm{e}^{i\omega_B t}dt=\mathrm{e}^{\frac{i\omega_B \tau}{2}}\mathrm{sinc} \Bigl(\frac{\omega_B \tau}{2}\Bigl)
	\end{equation}
	and $\tau$ is the coarse-graining time scale.
	
	Similarly, comparing the second part with eq.[(\ref{H_int}),(\ref{system ops})], we get  
	\begin{align*}
		& \lambda_{i}=\kappa , \quad p_{ii^{\prime}}(t)=\delta_{ii^{\prime}}\mathrm{e}^{i\omega_{0i}t},  \qquad q^{\star}(t)=\mathrm{e}^{-i\omega_B t}, \quad \hat{S}_{i^{\prime}}=\hat{a}_{i^{\prime}}     
	\end{align*}
 
	The corresponding dissipation parameter for the second part is 
	\begin{align}	
		\label{DP-2}
		\frac{\chi^{(2)}_{i^{\prime}j^{\prime}}(\tau)}{\tau}=& \tau \lambda^{\star}_{{i}^{\prime}} {\lambda}_{j^{\prime}}\; \langle {\hat{b}^{\dag}} \hat{b} \rangle_{B}\; \Gamma(\omega_{0i^{\prime}}-\omega_B) \Gamma(-\omega_{0j^{\prime}}+\omega_B)
	\end{align}
	where
	\begin{equation}
		\Gamma^{i^{\prime}}_{i}(\tau)
		=\delta_{ii^{\prime}}\; \Gamma(\omega_{0i}-\omega_B)
	\end{equation}
	
	The total dissipation parameter becomes from eq.(\ref{DP-1}) and eq.(\ref{DP-2}),
	\begin{equation}\label{DP_F}
		\begin{split}
			\gamma_{ij}		&=\frac{\chi^{(1)}_{ij}(\tau)}{\tau}+\frac{\chi^{(2)}_{ij}(\tau)}{\tau}\\&=\tau |\kappa |^{2}\biggl(\langle \hat{b}\hat{b}^{\dag}\rangle_{B}\; \Gamma(\omega_{0i}+\omega_B)\Gamma(-\omega_{0i}-\omega_B)+\langle \hat{b}^{\dag}\hat{b}\rangle_{B}\; \Gamma(\omega_{0i}-\omega_B)\Gamma(-\omega_{0i}+\omega_B)\biggl) \\
			&= \tau |\kappa|^{2}	\Biggl[	\Biggl( \frac{\mathrm{e}^{\beta\omega_B}}{\mathrm{e}^{\beta\omega_B}-1}	\Biggl) \Gamma(\omega_{0i}+\omega_B)	\Gamma(-\omega_{0j}-\omega_B)	+ \Biggl(\frac{1}{\mathrm{e}^{\beta\omega_B}-1}\Biggl)	\Gamma(\omega_{0i}-\omega_B)  \Gamma(-\omega_{0j}+\omega_B)	\Biggl]
		\end{split}
	\end{equation}
	The individual components are given by the explicit expressions
	\begin{align}
		\label{g11}
		\gamma_{11}		&=\tau |\kappa |^{2} \biggr[\biggl(\frac{\mathrm{e}^{\beta\omega_B}}{\mathrm{e}^{\beta\omega_B}-1}\biggl) \mathrm{sinc}^{2}\Bigl(\frac{\omega_{0}+\omega_B}{2}\tau\Bigl)     +      \biggl(\frac{1}{\mathrm{e}^{\beta\omega_B}-1}\biggl)  \mathrm{sinc}^{2}\Bigl(\frac{\omega_{0}-\omega_B}{2}\tau\Bigl)     \biggr]\\
		\label{g22}
		\gamma_{22}		&=\tau |\kappa |^{2} \biggr[\biggl(\frac{\mathrm{e}^{\beta\omega_B}}{\mathrm{e}^{\beta\omega_B}-1}\biggl) \mathrm{sinc}^{2}\Bigl(\frac{\omega_{0} - \omega_B}{2}\tau\Bigl)     +      \biggl(\frac{1}{\mathrm{e}^{\beta\omega_B}-1}\biggl)  \mathrm{sinc}^{2}\Bigl(\frac{\omega_{0} + \omega_B}{2}\tau\Bigl)     \biggr]\\
		\label{g12}
		\gamma_{12}		&=\tau |\kappa |^{2} \mathrm{sinc} \Bigl(\frac{\omega_{0}+\omega_B}{2}\tau\Bigl)  \mathrm{sinc} \Bigl(\frac{\omega_{0}-\omega_B}{2}\tau\Bigl)  \mathrm{coth}\Bigl( \frac{\beta\omega_B}{2} \Bigl) \mathrm{e}^{i\omega_{0} \tau}\\
		\label{g21}
		\gamma_{21}		&=\tau |\kappa |^{2} \mathrm{sinc}\Bigl(\frac{\omega_{0}+\omega_B}{2}\tau\Bigl) \mathrm{sinc}\Bigl(\frac{\omega_{0}-\omega_B}{2}\tau\Bigl)  \mathrm{coth}\Bigl( \frac{\beta\omega_B}{2} \Bigl) \mathrm{e}^{- i\omega_{0} \tau}
	\end{align} 
	The Lamb-shift Hamiltonian [eq.(\ref{lamb-para})]
	\begin{align} \label{lambexm}
		\hat{H}_{LS}&=\frac{i}{2}\sum_{\alpha}\langle \dot{\chi}_{\alpha 0}\rangle \hat{S}_{\alpha}-\langle \dot{\chi}_{\alpha 0}\rangle^{\star}\hat{S}^{\dag}_{\alpha}=0.
	\end{align}
	where [eq.(\ref{chi0})]
	\begin{equation}
		\begin{split}
			\langle \dot{\chi}_{\alpha 0}\rangle =&\frac{\chi_{\alpha 0}(\tau)}{\tau}
			=\sum_{i} \lambda_{i} \langle \hat{b} \rangle_{B}\;\Gamma^{i^{\prime}}_{i}(\tau)=0,~~~{\hbox{since}}~~~ \langle \hat{b} \rangle_{B}=0.
		\end{split}
	\end{equation}
	From eq.(\ref{system ops}) we can identify the the system operators in this model $\hat{S}_{1}=\hat{a}, \hat{S}_{2}=\hat{a}^{\dag}$. Now that both the dissipation parameters  [eq.(\ref{g11})-eq.(\ref{g21})] and the system operators have been identified we can test the correspondence explicitly in the next section.
	
	\subsection{The Quantum Side of the Correspondence} 
	The RHS of the equation stating the correspondence, eq.(\ref{correspondence}), which refers to the open quantum system becomes,
	\begin{align}\label{RHS1}
		\frac{1}{2}\gamma^{\alpha\beta}\hat{S}^{\dag}_{\beta}\hat{S}_{\alpha}=\frac{1}{2}\biggr[\gamma_{11}\;\hat{a}^{\dag}\hat{a}+\gamma_{22}\;\hat{a}\hat{a}^{\dag}+\gamma_{12}\;\hat{a}^{2}+\gamma_{21}\;(\hat{a}^{\dag})^{2}\biggr]    
	\end{align}
	The system's creation and annihilation operators can be written as,
	\begin{align*}
		\hat{a}=\frac{1}{\sqrt{2m\omega_{0}}}(m\omega_{0}\hat{x}_{1}+i\hat{p}_{1})\\
		\hat{a}^{\dag}=\frac{1}{\sqrt{2m\omega_{0}}}(m\omega_{0}\hat{x}_{1}-i\hat{p}_{1})
	\end{align*}
	Then eq.(\ref{RHS1}) becomes,
	\begin{equation}
		\begin{split}\label{RHS2}
			\frac{1}{2}\gamma^{\alpha\beta}\hat{S}^{\dag}_{\beta}\hat{S}_{\alpha}   &=   \frac{m\omega_0}{4} (\gamma_{11}+\gamma_{22}+\gamma_{12}+\gamma_{21})  \hat{x}_{1}^{2} \;  +  \frac{1}{4m\omega_0}(\gamma_{11}+\gamma_{22}-\gamma_{12}-\gamma_{21}) \; \hat{p}_{1}^{2}  \\
			&+ \frac{i}{4} (\gamma_{11}-\gamma_{22}+\gamma_{12}-\gamma_{21}) \; \hat{x}_1 \hat{p}_1   + \frac{i}{4} ( -\gamma_{11}+\gamma_{22}+\gamma_{12}-\gamma_{21}) \; \hat{p}_1 \hat{x}_1
		\end{split}
	\end{equation}
	Having calculated the RHS of eq.(\ref{correspondence}) that states the correspondence, all we need to do is to identify a suitable set of constraints on the phase space of the two oscillator system $(x_1, p_1,x_2,p_2)$ to realize the classical-quantum correspondence. We 
	tackle this issue in the next subsection.
	
	\subsection{The Classical Side of the Correspondence}
	Let us begin by imposing a constraint on the full phase space by setting a general linear combination of the phase space variables weakly to zero,
	\begin{equation}\label{p-constraint}
		\phi_{1}=\alpha x_{1}+\beta p_{1}+\gamma x_{2}+\delta p_{2} \approx 0.
	\end{equation}
	Thus, the primary Hamiltonian is given by,
	\begin{equation}
		\begin{split}
			H_{P}&=H_{C}+\lambda \phi_{1} \\
			&=\frac{1}{2}p^2_{1}+\frac{1}{2}k_{1}x^{2}_{1}+\frac{1}{2}p^2_{2}+\frac{1}{2}k_{2}x^{2}_{2}-k^{\prime}x_{1}x_{2}+\lambda (\alpha x_{1}+\beta p_{1}+\gamma x_{2}+\delta p_{2})
		\end{split}
	\end{equation}
	Requiring the constraint $\phi_1 \approx 0$, to be preserved in time,
	\[
	\dot{\phi_{1}}\overset{!}{\approx} 0
	\]
	implies,
	\begin{equation}
		\begin{split}
			\dot{\phi_{1}}   &\approx \Bigr\{\phi_{1},H_{P}\Bigr\}_P \approx 0\\
			&\Rightarrow (\beta k_{1}-\delta k^{\prime})x_{1}+(\delta k_{2}-\beta k^{\prime})x_{2}-\alpha p_{1}-\gamma p_{2} \not\approx 0
		\end{split}
	\end{equation}
	which gives us a secondary constraint,
	\begin{equation}
		\phi_{2}=(\beta k_{1}-\delta k^{\prime})x_{1}+(\delta k_{2}-\beta k^{\prime})x_{2}-\alpha p_{1}-\gamma p_{2}
	\end{equation}
	The secondary constraint should be consistent as well, \[
	\dot{\phi_{2}}\overset{!}{\approx} 0
	\]
	Thus, $\;\dot{\phi_{2}} \approx \Bigr\{\phi_{1},H_{P}\Bigr\}_P \approx 0$
	\begin{equation}
		\begin{split}
			\Rightarrow (k^{\prime}\gamma-k_{1}\alpha)x_{1}+(k^{\prime}\alpha-k_{2}\gamma)x_{2}&+(k^{\prime}\delta-k_{1}\beta)p_{1}+(k^{\prime}\beta-k_{2}\delta)p_{2}\\
			&+\lambda (-\alpha^{2}-k_{1}\beta^{2}-\gamma^{2}+2k^{\prime}\beta\delta-k_{2}\delta^{2}) \not\approx 0
		\end{split}
	\end{equation}
	We can solve for the Lagrange multiplier $\lambda$ 
	from the above equation which ensures the consistency of $\phi_{2}$.
	\begin{equation}
		\lambda=\frac{(k^{\prime}\gamma-k_{1}\alpha)x_{1}+(k^{\prime}\alpha-k_{2}\gamma)x_{2}+(k^{\prime}\delta-k_{1}\beta)p_{1}+(k^{\prime}\beta-k_{2}\delta)p_{2}}{\alpha^{2}+k_{1}\beta^{2}+\gamma^{2}-2k^{\prime}\beta\delta+k_{2}}
	\end{equation}
	In summary, we have two constraints in total and their Poisson bracket is a non-vanishing constant. So both the constraints are second class and we can construct the anti-symmetric matrix of Poisson brackets $C_{ab}$ and its inverse 
	$D_{ab}$ out of them as given below:
	
	\[C_{ab} \equiv \bigr\{ \phi_a,\phi_b \bigr\}_P \Rightarrow C =
	\begin{pmatrix}
		0 &\eta \\
		-\eta &0
	\end{pmatrix}
	\]
	\begin{equation}
		{\hbox{where}}, \;\; \eta = C_{12}=-C_{21}=\alpha^2+k_1\beta^2+\gamma^2+2k^\prime \beta\delta+k_2\delta^2
	\end{equation}
	So,
	\[D_{ab}=C^{-1}_{ab} \Rightarrow  D =
	\begin{pmatrix}
		0 & -\frac{1}{\eta} \\
		\frac{1}{\eta} &0
	\end{pmatrix}
	\]
	It follows that the classical side of the correspondence[eq.(\ref{correspondence})] is given by 
	\begin{equation}\label{LHS1}
		\begin{split}	
			D^{ab}\chi_b \phi_a   &=D^{ab}\; \dot{\phi_{b}}\phi_{a}\\
			&= D^{12} \dot{\phi_{2}}\phi_{1} +  D^{21} \dot{\phi_{1}}\phi_{2}   =\frac{1}{\eta}\phi^{2}_{2} \\
			&= \frac{1}{\eta} \biggr[(\beta ^2 k_{1}^{2}-2 \beta  \delta  k_1 k^{\prime}+\delta^2 {k^{\prime}}^2)x^{2}_{1}+(\beta ^2 {k^{\prime}}^2-2 \beta  \delta  {k^{\prime}}^{2} k_{2}+\delta ^2 k_{2}^{2})x^{2}_{2}+\alpha^{2}p^{2}_{1}+\gamma^{2}p^{2}_{2}\\
			&+(2 \alpha  \delta  k^{\prime}-2 \alpha  \beta  k_{1})x_{1}p_{1}+(2 \beta  \gamma  k^{\prime}-2 \gamma  \delta  k_{2})x_{2}p_{2}+(-2 \beta^2 k_{1} k^{\prime}+2 \beta  \delta  k_{1} k_{2}\\
			&+2 \beta  \delta  {k^{\prime}}^2-2 \delta^2 k^{\prime} k_{2})x_{1}x_{2}+2\alpha\gamma p_{1}p_{2}+(2 k^{\prime} \alpha \beta-2 k_2 \alpha \delta)x_{2}p_{1}+(-2 k_1 \beta \gamma+2 k^{\prime} \gamma \delta)x_{1}p_{2}\biggr]
		\end{split}
	\end{equation}
	where we have used $\chi_b = \dot\phi_b$.
	\subsection{The Classical-Quantum Correspondence}
	The classical-quantum correspondence follows from 
	consistency requirements. Comparing the two expressions for the quantum (eq.\ref{RHS2}) and the classical sides (eq.\ref{LHS1}) (using Weyl ordering for products of observables were necessary) evaluated above, it follows that 
	\begin{align}\label{modcoeff}
		\gamma=0,\quad\delta=\frac{k^{\prime}}{k_2}\beta
	\end{align}
	Then the constraints become
	\begin{align}
		\label{modphy1}
		\hat{\phi}_1 = \alpha \hat{x}_1 + \beta \Bigl(\hat{p}_1 + \frac{k^{\prime}}{k_2}\hat{p}_2 \Bigl) \\
		\label{modphy2}
		\hat{\phi}_2 = \bigl(k_1 -\frac{{k^{\prime}}^2}{k_2}\bigl)\beta \hat{x}_1 -\alpha \hat{p}_1 
	\end{align}
	and eq.(\ref{LHS1}) becomes,
	\begin{align}\label{LHS2}
		D^{ab} \hat{\chi}_b \hat{\phi}_a =	\frac{1}{\eta}		       \Biggr[\frac{({k^{\prime}}^2-k_{1}k_{2})^{2}\beta^{2}}{k_{2}^{2}}\hat{x}^{2}_{1}  +  \alpha^{2}\hat{p}^{2}_{1}  + \frac{{k^{\prime}}^2 -k_1 k_2}{k_2} \alpha\beta\; (\hat{x}_1 \hat{p}_1 + \hat{p}_1 \hat{x}_1) \Biggr]
	\end{align}
	with
	\begin{equation}\label{eta}
		\eta = \alpha^{2}+\Bigl( k_1 - \frac{{k^{\prime}}^2}{k_2}\Bigl) \beta^2
	\end{equation}
	Finally, comparing the coefficients of $x^{2}_{1}$,$\;p^{2}_{1}$,$\; x_{1}p_{1}$ and $\; p_1 x_1 $ [eq.(\ref{LHS2})] with the RHS [eq.(\ref{RHS2})], we get
	\begin{align}
		\label{c1}
		\frac{m\omega_0}{4} (\gamma_{11}+\gamma_{22}+\gamma_{12}+\gamma_{21}) &=  \frac{1}{\eta}   \frac{({k^{\prime}}^2-k_{1}k_{2})^{2}\beta^{2}}{k_{2}^{2}} \\
		\label{c2}
		\frac{1}{4m\omega_0}(\gamma_{11}+\gamma_{22}-\gamma_{12}-\gamma_{21})  &=   \frac{1}{\eta}  \alpha^2 \\
		\label{c3}
		\frac{i}{4} (\gamma_{11}-\gamma_{22}+\gamma_{12}-\gamma_{21})         &=   \frac{1}{\eta}   \frac{{k^{\prime}}^2 -k_1 k_2}{k_2} \alpha\beta   \\
		\label{c4}
		\frac{i}{4} ( -\gamma_{11}+\gamma_{22}+\gamma_{12}-\gamma_{21})       &=   \frac{1}{\eta}   \frac{{k^{\prime}}^2 -k_1 k_2}{k_2} \alpha\beta
	\end{align}
	These equations relate the dissipation parameters in the reduced quantum system to the coefficients of the phase space variables in the constraints that define the classical constrained system. We need to solve these equations for $\alpha,\beta$ in order to determine the constraints [eq.(\ref{modphy1}),(\ref{modphy2})].
	
	Note that eq.(\ref{c3}) and  eq.(\ref{c4}) are not consistent unless $\; \gamma_{11} = \gamma_{22}$. Therefore equating the expressions for $\gamma_{11}$ and $\gamma_{22}$ in eq.(\ref{g11}) and eq.(\ref{g22}) leads to the following 
	\begin{align}\label{sinc}
		\mathrm{sinc}\Big(\frac{\omega_0 +\omega_B}{2}\tau\Bigl) = \pm \; \mathrm{sinc}\Bigl(\frac{\omega_0 -\omega_B}{2}\tau\Bigl)
	\end{align}
	which has two regimes {\it viz.} $\omega_B \gg \omega_0$ or $\omega_0 \gg \omega_B$. The Markovian approximation suggests the bath characteristic
	timescale has to be much faster than the system’s, that is $\omega_B^{-1}<<\omega_0^{-1}$. Thus it is sensible to choose $\omega_B \gg \omega_0$.
	
	In this limit, the dissipation parameters [eq.(\ref{g11}) - eq.(\ref{g21})] become,
	\begin{equation}
		\begin{split}\label{gamma11}
			\gamma_{11} =  \gamma_{22}  =  \tau |\kappa|^2 \mathrm{coth}\Bigl(\frac{\beta\omega_B}{2}\Bigl) \mathrm{sinc}^2(\frac{\omega_B\tau}{2})~~~ 
			\hbox{and,\;\;}\\\gamma_{12}  =  \gamma_{11} \mathrm{e}^{i\omega_{0}\tau}  \qquad   \gamma_{21}  =  \gamma_{11} \mathrm{e}^{-i\omega_{0}\tau}
		\end{split}
	\end{equation}
	and the four equations (\ref{c1} - \ref{c4}) become,
	\begin{align}
		\label{c5}
		m\omega_{0}\gamma_{11} \mathrm{cos}^2\Bigl(\frac{\omega_{0}\tau}{2}\Bigl)  &=  \frac{1}{\eta} \frac{({k^{\prime}}^2-k_1 k_2)^2}{k_{2}^{2}} \beta^2  \\
		\label{c6}
		\frac{1}{m\omega_0} \gamma_{11} \mathrm{sin}^2\Bigl(\frac{\omega_{0}\tau}{2}\Bigl)   &= \frac{1}{\eta} \alpha^2  \\
		\label{c7}
		-\frac{\gamma_{11}}{2} \mathrm{sin}(\omega_{0} \tau) &=   \frac{1}{\eta} \frac{{k^{\prime}}^2-k_1 k_2}{k_{2}} \alpha\beta 
	\end{align}
	Note that multiplying eq.(\ref{c5}) and eq.(\ref{c6}) we get the square of eq.(\ref{c7}),
	\begin{equation}
		\frac{\gamma_{11}^2}{4}   \mathrm{sin}^2 (\omega_{0}\tau )  =   \Bigl( \frac{{k^{\prime}}^2-k_1 k_2}{k_{2}}\Bigl)^2  \;\; \frac{\beta^2 \alpha^2}{\eta^2}
	\end{equation}
	It therefore suffices to solve the independent equations (\ref{c5}) and (\ref{c6}) for $\alpha$ and $\beta$. With $\eta$ given by eq.(\ref{eta}), we can compute $\alpha$ in terms of $\beta$ from eq.(\ref{c5}),
	\begin{equation}
		\label{alp1}
		\alpha = \pm \Biggl( \frac{{k^{\prime}}^2-k_1 k_2}{k_2} \Bigl[  1+\frac{1}{c_0} \frac{{k^{\prime}}^2-k_1 k_2}{k_2} \Bigl] \Biggl)^{\frac{1}{2}} \beta
	\end{equation}
	where
	\begin{equation}
		c_0 = m\omega_{0}\gamma_{11} \mathrm{cos}^2\Bigl(\frac{\omega_{0}\tau}{2}\Bigl) 
	\end{equation}
	Similarly from eq.(\ref{c6}) we get,
	\begin{equation}
		\label{alp2}
		\alpha = \pm \Biggl( \frac{c_1}{c_1 - 1} \frac{{k^{\prime}}^2-k_1 k_2}{k_2} \Biggl)^{\frac{1}{2}}\beta
	\end{equation}
	where
	\begin{equation}
		c_1 = \frac{1}{m\omega_0} \gamma_{11} \mathrm{sin}^2\Bigl(\frac{\omega_{0}\tau}{2}\Bigl)
	\end{equation}
	So the constraints eq.(\ref{modphy1}) and eq.(\ref{modphy2}) become,
	\begin{equation}\label{P2}
		\hat{\phi}_1 = \alpha \hat{x}_1 + \beta \Bigl(\hat{p}_1 + \frac{k^{\prime}}{k_2}\hat{p}_2 \Bigl)
	\end{equation}
	\begin{equation}
		\hat{\phi}_2 = \bigl(k_1 -\frac{{k^{\prime}}^2}{k_2}\bigl)\beta \hat{x}_1 -\alpha \hat{p}_1 
	\end{equation}
	where, $\alpha$ is given by eq.(\ref{alp1}) and eq.(\ref{alp2}). Choosing $\beta = 1$ without loss of generality, we find four possible values of $\alpha$
	and hence four ways of constructing the constraint
	$\hat{\phi}_1$.\\
	
	The last remaining piece of the jigsaw puzzle consists in identifying the corresponding Lindblad operators. 
	
	This is easily done by diagonalizing the dissipation matrix via eq.(\ref{diag_gamma})
	\begin{equation}\label{diag_gamma_matrix}
		\tilde{\gamma} = 
		\begin{pmatrix} 
			2\gamma_{11} & 0 \\
			0   & 0 
		\end{pmatrix}
	\end{equation}
	with,
	\begin{equation*}
		u = \frac{1}{\sqrt{2}\gamma_{11}}
		\begin{pmatrix}
			\gamma_{12} & -\gamma_{12} \\
			\gamma_{11} & \gamma_{11}
		\end{pmatrix}
	\end{equation*}
	The Lindblad operators can be found from eq.(\ref{lindbladops})
	\begin{equation}
		\hat{L}_1 =\frac{1}{\sqrt{2}} \Bigl(\mathrm{e}^{i\omega_0 \tau} \hat{a} + \hat{a}^{\dag}\Bigl); \qquad  \hat{L}_2 = \frac{1}{\sqrt{2}} \Bigl(-\mathrm{e}^{i\omega_0 \tau} \hat{a} + \hat{a}^{\dag}\Bigl)
	\end{equation}
	where,
	\begin{equation*}
		\hat{S} =
		\begin{pmatrix}
			\hat{a},  & \hat{a}^{\dag}
		\end{pmatrix}
	\end{equation*}
	From eq.(\ref{main c}) we can get the Lindblad operators in terms of the constraints
	\begin{equation}\label{lind_dissi}
		\hat{L}_1 \hat{L}^{\dag}_1 = \frac{1}{2\gamma_{11}\eta} \hat{\phi}^{2}_{2} 
	\end{equation}
	where, $\hat{\phi}_2$, $\eta$ and $\gamma_{11}$ are given by eq.(\ref{modphy2}), eq.(\ref{eta}) and eq.(\ref{gamma11}) respectively. Note that $\hat{L}_2$ doesn't appear in the above equation because the corresponding element in the diagonalized dissipation matrix [eq.(\ref{diag_gamma_matrix})]  $\tilde{\gamma}_{22} = 0$.

	\section{Conclusions and Outlook}\label{S4}
	In this paper, we have explored an intriguing 
	connection between constrained classical mechanical systems on the one hand, and open quantum systems on the other hand. We derived a precise connection between the two by comparing the generalizations of the Liouville equation which describes the time evolution of the system. The Lindblad operators which capture the entanglement between the system and the bath in the quantum case, and are responsible for dissipation, decoherence, etc., are mapped to the constraint equations in the classical case which are responsible for the reduction of the classical phase space, and hence to a modified time evolution equation involving Dirac brackets. 
	
	We illustrated this novel correspondence in a
	concrete example, namely coupled harmonic oscillators.  
	
	Although the correspondence we established may seem puzzling at first sight, it is no more surprising than the many well-established connections between quantum mechanics and classical mechanics {\it viz.} Bohr's Correspondence Principle, the Ehrenfest Theorem, the JWKB approximation, Coherent States, and the Wigner distribution.
	
	Our results shed new light on both open quantum systems and constrained classical systems. Further investigations along these lines should enable us to study gauge theories from an open quantum systems perspective and to understand open quantum systems from a geometric and gauge theoretic perspective.  We hope to report on these developments in the near future.

	\section{Acknowledgement}
	This work is partially funded by a grant from the Infosys Foundation. We thank the referee for insightful comments which forced us to considerably improve the presentation of the paper. We also thank H S Mani for discussions.

	\appendix
	\section{Constrained Dynamical Systems}\label{DCSA}
	In this appendix we will give a crash tutorial on Constrained Dynamical Systems  \cite {dirac}\cite{Weinberg}\cite{Hanson}\cite{Mukunda} \cite{KS}\cite{Henneaux}.
	
	Let $S$ denote the action functional
	\begin{equation}
		S[q_{i}(t)]=\int_{a}^{b} dt~L(q_{i},\dot{q}_{i})
	\end{equation}
	where $q_{i}(t)$ are canonical coordinates and $\dot{q}_{i}(t)$ denotes derivative of 
	$q_i$ with respect to time $t$. For simplicity, we consider Lagrangians without explicit time dependence. \\
	
	The Poisson brackets between the coordinates and the canonically conjugate momenta $p_i = {\partial L\over \partial \dot q_i}$ are given by the equation
	\begin{equation}
		\Bigl\{q_{j},p_{i}\Bigl\}_P = \delta_{ij}
	\end{equation}
	The canonical Hamiltonian is obtained by a 
	Legendre transform of the Lagrangian in the 
	following manner,
	\begin{equation}
		H_{c}=p_{i}\dot{q}_{i}-L(q_{i},\dot{q}_{i})
	\end{equation}
	The dynamics is determined by Hamilton's equations
	\begin{equation}
		\dot{q}_{i}=\Bigl\{q_{i},H_{c}\Bigl\}_P=\frac{\partial H_{c}}{\partial p_{i}}, \qquad \dot{p}_{i}=\Bigl\{p_{i},H_{c}\Bigl\}_P= - \frac{\partial H_{c}}{\partial q_{i}}
	\end{equation}
	If the Lagrangian $L(q_{i},\dot{q}_{i})$ is singular, the determinant of the Hessian $W_{ij}$ vanishes,
	\begin{equation}\label{hessian}
		Det \Bigr[W_{ij}\Bigr]=0 \qquad {\hbox{where}}~ W_{ij}=\frac{\partial^{2}L}{\partial \dot{q}_{i}\partial \dot{q}_{j}}  =  \frac{\partial p_{i}}{\partial \dot{q}_{j}}
	\end{equation} 
	If the rank of the Hessian $W_{N\times N}$ is R $(R<N)$ and $2N$ is the total number of phase space variables,
	\begin{equation}\label{qdota}
		\dot{q}^{a}=f^{a}(q,p_{b},\dot{q}^{\rho})\qquad a, b =1,..,R \qquad {\hbox{and}} \qquad \rho =R+1,...,N
	\end{equation}
	From eq.(\ref{hessian}) and eq.(\ref{qdota}) we can see that $(N - R)$ $\dot{q}_{i}$ are undetermined,
	\begin{align}
		p_{i}&=\tilde{g}_{i}(q,\dot{q}^{a},\dot{q}^{\rho})=\tilde{g}_{i}\Bigl(q,f^{a}(q,p_{b},\dot{q}^{\rho}),\dot{q}^{\rho}\Bigl)  =  g_{i}(q,p_{b},\dot{q}^{\rho})   \quad i=1,2...N\\
		\label{constraint}
		p_{\tau}&=g_{\tau}(q,p_{b})\qquad \tau =R+1,...,N
	\end{align}
	The $p_{\tau}$ does not depend on $\dot{q}^{\rho}$s since we can't solve for them.
	These $(N - R)$ relations given by eq.(\ref{constraint}) are the primary constraints,
	\begin{equation}\label{P-constraint}
		\phi_{m}(q,p)\approx 0. \qquad {\hbox{ where}}, m =R+1,...,N
	\end{equation}
	The `$\approx$' sign, said to be weakly equal to, indicates that this equation identically holds on the subspace determined by the constraints, but not on the whole phase space. 
	
	The canonical Hamiltonian is restricted to the reduced phase space and is not uniquely determined. We may add to it any linear combination of the $\phi_m$ which will give us the primary Hamiltonian.
	\begin{equation} 
		H_{p}=H_{c}+\lambda_{m} \phi_{m} \approx H_c
	\end{equation}
	where the $\lambda_m$ are the Lagrange multipliers. The requirement that the primary constraints [eq. (\ref{P-constraint})] be preserved in time yields,
	\begin{equation}
		\dot{\phi}_{n}=\Bigl\{\phi_{n},H_{p}\Bigl\}_P \approx \Bigl\{\phi_{n},H_{c}\Bigl\}_P+\lambda_{m}\Bigl\{\phi_{n},\phi_{m}\Bigl\}_P \approx 0. \;\; \forall n=R+1...N
	\end{equation}
	Three things can happen at this stage:\\ 
	(a) the above equation can be identically satisfied, in which case
	no extra information can be gathered about the system,\\
	(b) the above gives rise to a new equation depending on the Lagrange multipliers $\lambda_m$, using which we can determine one or more
	of them explicitly as functions of the coordinates and momenta,\\
	(c) we get a new equation between the coordinates and momenta, independent of the Lagrange multipliers, $\lambda_m$. These new relations are also constraints.
	
	We need to go through this rigmarole once again with the new set of constraints. At the end of this exercise, we would have determined some $\lambda_m$, and the complete set of new constraints we acquire in the process is called the set of secondary constraints.
	These, together with the primary constraints form the total set of constraints. To summarise,
	\begin{equation}
		\phi_{a}(q,p)\approx 0  \qquad a=1,...T,\;\; {\hbox{where}},\; T=K+M\;{\hbox{and}}\; M=N-R
	\end{equation}
	In the above equation, $M$ is the number of primary constraints, $K$ is the number of additional secondary constraints, and $T$ is the total number of constraints. The remaining arbitrariness in the Hamiltonian is due to the Lagrange multipliers that are left undetermined at this stage.
	
	Dirac defines a function $R(q,p)$ as a first-class quantity if
	\begin{equation}
		\Bigl\{R,\phi_{a}\Bigl\}_P ~\approx 0, \qquad a=1,...,T
	\end{equation}
	and second class if
	\begin{equation}\label{2nd class}
		\Bigl\{R,\phi_{a}\Bigl\}_P ~\not\approx 0.\quad \text{for at least one $a$.}
	\end{equation}
	All the constraints can be classified into first and second-class constraints:
	\begin{align}
		\psi_{i}(q,p)&\approx 0.\quad i=1,...,I \quad\text{[First-class constraints]}\\
		\phi_{\alpha}(q,p)&\approx 0. \quad \alpha=1,...,Q \quad \text{[Second-class constraints]}
	\end{align}
	First-class constraints generate redundancies (gauge invariances) in theory. By choosing an appropriate number of gauge fixing conditions, these redundancies can be eliminated, and the first-class constraints are converted to second-class by the gauge fixing conditions. One may check that these extra gauge fixing constraints 
	are precisely the ones that fix the remaining undetermined Lagrange multipliers. 
	
	The Dirac bracket between any two observables $A$ and $B$, is defined as follows:
	\begin{align}
		\Bigl\{A,B\Bigl\}^{\star}=\Bigl\{A,B\Bigl\}_P-\Bigl\{A,\phi_{\alpha}\Bigl\}_P C^{-1}_{\alpha\beta}\Bigl\{\phi_{\beta},B\Bigl\}_P . \qquad 
	\end{align}
	where $\phi_\alpha$ are second class constraints, and $C_{\alpha\beta}\equiv \Bigl\{\phi_{\alpha},\phi_{\beta}\Bigl\}_P$ are the components of the matrix of Poisson brackets of second class constraints.
	
	The time development of any classical observable, $R(q,p)$ is determined by the 
	evolution equation
	\begin{equation}
		\dot{R}(t)=\bigr\{R,H_c \bigr\}^{\star} = \bigr\{R ,H_c  \bigr\}_P -\bigr\{R,\phi_{\alpha}\bigr\}_P ~ C^{-1}_{\alpha\beta} ~ \bigr\{\phi_{\beta},H_c \bigr\}_P 
	\end{equation}
	The evolution equation is exactly like the Liouville equation but with the Poisson bracket replaced by the Dirac bracket.

	\section{Lindblad Equation} \label{LindbladB}
	Lindblad, and independently Gorini, Kossakowski, and Sudarshan,  derived the most general completely positive Markovian semigroup master equation for the dynamics of an open quantum system \cite{L original}\cite{GKS original}\cite{Manzano} given by,
	\begin{equation} \label{lindblad}
		\dot{\hat{\rho}}(t) = -i [\hat{H}_s , \hat{\rho}(t)] + \sum_{\alpha,\beta} \gamma_{\alpha\beta} \Bigl[\hat{S}_{\alpha}\hat{\rho}(t)\hat{S}^{\dag}_{\beta}-\frac{1}{2}\bigr\{\hat{S}^{\dag}_{\beta}\hat{S}_{\alpha},\hat{\rho}(t) \bigr\} \Bigl]
	\end{equation}
	In this equation $\hat{S}_\alpha, ~\alpha = 1,2,\cdots N^2$, where $N$ is the dimension of system Hilbert space $\mathcal{H_S}$, stand for an orthonormal set of operators, and $\gamma$ is a constant positive semi-definite matrix of 
	dissipation parameters. The orthonormality of the set of operators can be verified
	with the Hilbert-Schmidt inner product, $\braket{\braket{\hat{S}_{\alpha}|\hat{S}_{\beta}}}\equiv Tr(\hat{S}^{\dag}_{\alpha}\hat{S}_{\beta})=\delta_{\alpha ,\beta}$ \cite{Lspace}. In the above equation and in what follows, we set $\hbar=1$.
	
	In this section we review the master equation by starting from the full system, {\it i.e.}, the 
	bath, and the system, and derive the reduced density matrix by 
	integrating out the bath degrees of freedom. In doing so, we 
	explicitly identify the role played by the Markovian approximation via the coarse-graining procedure, which will enable us to derive 
	the all-important dissipation parameters. \\
	
	Let us consider a general Hamiltonian of the form,
	\begin{equation}
		\hat{H} =\hat{H}_{S}+\hat{H}_{B}+\hat{H}_{I}=\hat{H}_{0}+\hat{H}_{I}
	\end{equation}
	where, $\hat{H}_{0}=\hat{H}_{S}+\hat{H}_{B}$ is the `free' Hamiltonian of
	both the system and the bath, and $\hat{H}_{I}$ is the interaction Hamiltonian. Let us further assume initial decoupling between the system and bath,
	\begin{equation}
		\hat{\rho}_{T}(0)=\hat{\rho}_{S}(0)\otimes \hat{\rho}_{B}(0)
	\end{equation}
	
	The integration over the bath degrees of freedom can be performed in the usual way \cite{breuer}\cite{Rivas}\cite{lidar}. The time dependence of the reduced
	density operator can then be described by the Kraus Operator Sum 
	Representation (OSR), as follows:
	\begin{equation}
		\hat{\rho}_{s}(t)  =\sum_{lm}\hat{K}_{lm}(t)\hat{\rho}_{s}(0)\hat{K}^{\dag}_{lm}(t)  
	\end{equation}
	The Kraus operators straddle the total system (bath + system), and are matrix elements in so far as the bath is concerned, but act as operators on the system
	\begin{equation}\label{kraus}
		\hat{K}_{lm}(t)=\sqrt{p_{m}}\bra{l}\hat{U}(t)\ket{m}, \qquad \hat{U}(t)=\mathrm{e}^{-i\hat{H}t}
	\end{equation}
	The $\bigl\{ \ket{m} \bigl\}$ are the orthonormal eigenbasis of $\hat{\rho}_{B}(0)$ with non-negative eigenvalues $p_m$. Thus $l$ and $m$ can take values $0,1,2...d_B -1$.
	Evidently, there are $d_B^2$ number of Kraus operators, where
	$d_B$ is the dimensionality of the bath Hilbert space $\mathcal{H_B}$. We label them by $i = 0,1,2, \cdots d_B^2 -1$. 
	
	The Kraus operators, $\hat{K}_i$, can be expanded in the time-independent operator basis 
	$\Bigl\{\hat{S}_\alpha \Bigl\}^{d^{2}_{S}-1}_{\alpha =0}$ with $\hat{S}_0 = \hat{I}$,
	\begin{equation} \label{kra}
		\hat{K}_{i}(t)=\sum_{\alpha =0}^{d^{2}_{S}-1} b_{i \alpha}(t) \hat{S}_\alpha
	\end{equation}
	where, $b_{i\alpha}$ are the time-dependent elements of a rectangular $d^{2}_{S}\times d^{2}_{B}$ dimensional matrix.\\
	The reduced density operator can be expressed as
	\begin{equation} \label{rho-sys}
		\hat{\rho}_{s}(t)=\sum_{\alpha,\beta =0}^{d^{2}_{S}-1}\chi_{\alpha\beta}(t)\hat{S}_{\alpha}\hat{\rho}_{S}(0)\hat{S}^{\dag}_{\beta}.
	\end{equation}
	Note that the matrix elements $\chi_{\alpha\beta}(t)$ appearing in the above equation are given by
	\begin{equation} \label{chi}
		\chi_{\alpha\beta}(t)=\sum_{i=0}^{d_{B}^{2}-1} b_{i\alpha}(t)b^{\star}_{i\beta}(t)
	\end{equation}
	$\chi(t)$ is a positive semi-definite, hermitian $d^{2}_{s} \times d^{2}_{s}$ matrix. \\
	After some algebraic manipulations, we get the Fixed-Basis Operator Sum Representation Equation\cite{CG original}\cite{lidar}\cite{LBW} to be
	\begin{equation}\label{F-OSR}
		\frac{d}{dt}\hat{\rho}_{s}(t)=-i\Bigr[\dot{\hat{Q}}(t),\hat{\rho}_{s}(0)\Bigr]+\sum_{\alpha,\beta \geq 1}\dot{\chi}_{\alpha\beta}(t)\biggl(\hat{S}_{\alpha}\hat{\rho}_{s}(0)\hat{S}^{\dag}_{\beta}-\frac{1}{2}\Bigl\{\hat{S}^{\dag}_{\beta}\hat{S}_{\alpha},\hat{\rho}_{s}(0)\Bigl\}\biggl)
	\end{equation}
	where $\hat{Q}(t)$, whose physical significance will become clear presently, is a hermitian operator defined by,
	\begin{equation}
		\hat{Q}(t)\equiv \frac{i}{2} \sum_{\beta\geq 1} \Bigl(\chi_{\beta 0}(t)\hat{S}_{\beta} -\chi_{0\beta}(t)\hat{S}^{\dag}_{\beta} \Bigl)
	\end{equation}
	This equation connects the reduced density matrix at two different instants of time. To bring it to the form [eq.(\ref{CG-SP})], we resort to 
	the following procedure.
	
	\subsubsection*{Coarse Graining}
	The method of coarse-graining can be found in \cite{lidar}\cite{CG original}\cite{LBW}. We consider three time scales: (a) $\tau_{B}$, the bath time scale, typically the inverse of the maximum natural frequency associated with the bath (b) the coarse-graining time scale $\tau$ over which the bath is reset to its original state (timescale for the bath memory to be erased, also referred to as the Markovian approximation) and (c) a system timescale  
	$\tau_{s}$, the timescale associated with the changes
	in the reduced density matrix. The coarse-graining procedure consists of assuming
	\begin{equation}\label{CG_limit}
		\tau_{B} \ll \tau \ll \tau_{s}	
	\end{equation}
	Note that there is yet another timescale $\tau_0$ associated with the 
	unitary evolution of the system, and it should lie between $\tau_B$ and $\tau_s$.
	To quote Feynman, we are therefore in an `equilibrium' regime where "all the fast processes have already 
	taken place, and the slow processes are yet to take
	place"\cite{Feynman}.   
	
	Under these conditions, the Fixed-Basis OSR evolution equation(\ref{F-OSR}) reduces to 
	\begin{equation}
		\frac{d}{dt}\hat{\rho}_{s}(t)=-i\Bigr[\langle\dot{\hat{Q}}\rangle,\hat{\rho}_{s}(t)\Bigr]+\sum_{\alpha,\beta \geq 1}\langle\dot{\chi}_{\alpha\beta}\rangle\biggl(\hat{S}_{\alpha}\hat{\rho}_{s}(t)\hat{S}^{\dag}_{\beta}-\frac{1}{2}\Bigl\{\hat{S}^{\dag}_{\beta}\hat{S}_{\alpha},\hat{\rho}_{s}(t)\Bigl\}\biggl)
	\end{equation}
	where $\langle{X}\rangle=\frac{1}{\tau}\int_{0}^{\tau}X(t)dt$. $\langle\dot{\hat{Q}}\rangle$ is effectively the Hamiltonian responsible for the unitary evolution of the system: it contains not just the system Hamiltonian, but also the Lamb shift correction coming from the system-bath interaction. $\langle\dot{\chi}_{\alpha\beta}\rangle$ are the dissipation parameters. Thus, we have 
	\begin{equation} \label{CG}
		\frac{d}{dt}\hat{\rho}_{s}(t)=-i\Bigr[\hat{H}_{S}+\hat{H}_{LS},\hat{\rho}_{s}(t)\Bigr]+\sum_{\alpha,\beta \geq 1}\langle\dot{\chi}_{\alpha\beta}\rangle\biggl(\hat{S}_{\alpha}\hat{\rho}_{s}(t)\hat{S}^{\dag}_{\beta}-\frac{1}{2}\Bigl\{\hat{S}^{\dag}_{\beta}\hat{S}_{\alpha},\hat{\rho}_{s}(t)\Bigl\}\biggl).
	\end{equation}
	where $\langle\dot{\chi}_{\alpha\beta}\rangle=\gamma_{\alpha\beta}$[eq.(\ref{CG-SP})].
	
	\subsubsection*{Dissipation Parameters}
	In the interaction picture, the Hamiltonian becomes
	\begin{equation}
		\hat{\Tilde{H_{I}}}(t)=\hat{U}_{0}^{\dag}(t)\hat{H}(t)\hat{U}_{0}(t), ~~\qquad \hat{U}_{0}(t)=\mathrm{e}^{-i\hat{H}_{S}t}\otimes \mathrm{e}^{-i\hat{H}_{B}t}
	\end{equation}
	We take the interaction Hamiltonian to be of the form
	\begin{equation}
		\hat{\tilde{{H}}}_I(t) = \sum_{\beta,\delta} \lambda_{\beta \delta}\; \hat{\tilde{S}}_{\beta}(t) \otimes \hat{\tilde{B}}_{\delta}(t)
	\end{equation}
	where, $\hat{\tilde{S}}_{\beta}(t)$ and $\hat{\tilde{B}}_{\delta}(t)$ are the system and bath operators respectively, and can be written as
	\begin{equation}
		\begin{split}
			\hat{\tilde{S}}_{\beta}(t)=\mathrm{e}^{i\hat{H}_{S}t}\hat{S}_{\beta}\mathrm{e}^{-i\hat{H}_{S}t}= \sum_{\alpha} p_{\beta\alpha}(t)\hat{S}_{\alpha} \\
			\hat{\tilde{B}}_{\delta}(t)=\mathrm{e}^{i\hat{H}_{B}t}\hat{B}_{\delta}\mathrm{e}^{-i\hat{H}_{B}t}= \sum_{\gamma} q_{\delta\gamma}(t)\hat{B}_{\gamma}
		\end{split}
	\end{equation}
	The density matrix in the interaction picture evolves as
	\begin{equation}
		\hat{\tilde{\rho_{T}}}(t)=\hat{U}^{\dag}_{0}(t)\hat{\rho}_{T}(t)\hat{U}_{0}(t)=\hat{\tilde{U}}(t)\hat{\rho}_{T}(0)		\hat{\tilde{U}}^{\dag}(t)
	\end{equation}
	where $\hat{\rho}_{T}$ is the total density matrix of the system and bath collectively. The unitary evolution operator in the interaction picture is given by,
	\begin{equation}\label{UNI-IP}
		\begin{split}
			\hat{\tilde{U}}(t)&=\hat{U}^{\dag}_{0}(t)\hat{U}(t,0)\\
			&=\hat{T}\mathrm{exp}\biggr[-i\int_{0}^{t}\hat{\tilde{H}}_{I}(t^{\prime})dt^{\prime}\biggr]\\
			&=\hat{I}+\sum_{n=1}^{\infty}\frac{(-i)^{n}}{n!} \int_{0}^{t}dt_{n}\int_{0}^{t_{n}}dt_{n-1}....\int_{0}^{t_{2}}dt_{1} \hat{T}\Bigl\{\hat{\tilde{H}}_{I}(t_{1})\hat{\tilde{H}}_{I}(t_{2})....\hat{\tilde{H}}_{I}(t_{n})\Bigl\}
		\end{split}
	\end{equation}
	where $\hat{T}$ is the time ordering operator.
	
	The system density matrix in the interaction picture is,
	\begin{equation}
		\hat{\tilde{\rho_{S}}}(t)=\mathrm{Tr}_{B}\Bigr[\hat{\tilde{\rho_{T}}}(t)\Bigr]=\sum_{i} \hat{\tilde{K}}_{i}(t)\hat{\rho}_{S}(0)\hat{\tilde{K}}^{\dag}_{i}(t)
	\end{equation}
	where, $\tilde{K}_{i}$ are the Kraus operators in the interaction picture
	\begin{equation}\label{kraus_int}
		\hat{\tilde{K}}_{i}(t)=\hat{\tilde{K}}_{lm}(t)=\sqrt{p_{m}}\bra{l}\hat{\tilde{U}}(t)\ket{m}
	\end{equation}
	Then equations (\ref{kra}), (\ref{rho-sys}) and (\ref{chi}) become,
	\begin{equation}\label{int-chi}
		\hat{\tilde{K}}_{i}(t)=\sum_{\alpha =0}^{d^{2}_{S}-1} \tilde{b}_{i\alpha}(t) \hat{S}_{\alpha}; \qquad \hat{\tilde{\rho_{S}}}(t)=\sum_{\alpha,\beta =0}^{d^{2}_{S}-1} \tilde{\chi}_{\alpha\beta}(t)\hat{S}_{\alpha}\hat{\rho}_{S}(0)\hat{S}^{\dag}_{\beta} \qquad \tilde{\chi}_{\alpha\beta}(t)=\sum_{i=0}^{d_{B}^{2}-1}\tilde{b}_{i\alpha}(t)\tilde{b}^{\star}_{i\beta}(t)
	\end{equation}

	The explicit derivation of dissipation parameters can be found in \cite{LBW}.
	We can expand the Kraus operators by using the expansion of $\hat{\tilde{U}}(t)$ eq.(\ref{UNI-IP}). From now on we suppress the tilde notation for convenience. From eq.(\ref{kraus_int}) and eq.(\ref{UNI-IP}) with $\hat{S}_0 =\hat{I}_S$
	\begin{equation}
		\hat{K}_{i}(t)=\sqrt{p_{m}}\delta_{lm}\hat{I}_{S}+\sum_{n=1}^{\infty}\hat{K}^{(n)}_{i}(t)
	\end{equation}
	In the weak coupling regime, we will only consider up to the first  order ($n=1$) and we have,
	\begin{equation}\label{KK}
		\hat{K}^{(1)}_{i\equiv lm}(t)=-it\sqrt{p_{m}}\sum_{\alpha; \gamma; \beta, \delta} \lambda_{\beta\delta} \hat{S}_{\alpha} \bra{l}\hat{B}_{\gamma}\ket{m}\Gamma^{\alpha\gamma}_{\beta\delta} \equiv \sum_{\alpha =1}^{d^{2}_{S}-1}b_{i\alpha}(t)\hat{S}_{\alpha}
	\end{equation}
	where we define,
	\begin{equation}\label{gam}
		\Gamma^{\alpha\gamma}_{\beta\delta}(t)=\frac{1}{t}\int_{0}^{t} dt^{\prime} p_{\beta\alpha}(t^{\prime})q_{\delta\gamma}(t^{\prime})
	\end{equation}
	From eq.(\ref{KK}) and eq.(\ref{gam}) we can find that,
	\begin{equation}
		\begin{split}
			b_{i\alpha}(t)=-it\sqrt{p_{m}}\sum_{\alpha^{\prime} \alpha^{\prime\prime} \alpha^{\prime\prime\prime}}  
			\lambda_{\alpha^{\prime\prime} \alpha^{\prime\prime\prime}}  \bra{l}\hat{B}_{\alpha^{\prime}}\ket{m} \Gamma^{\alpha \alpha^{\prime}}_{\alpha^{\prime\prime} \alpha^{\prime\prime\prime}}(t) \qquad \alpha \geq 1
		\end{split}
	\end{equation}
	
	Using eq.(\ref{int-chi}),
	\begin{equation}\label{chi0}
		\chi_{\alpha 0}(t)=-it \sum_{\alpha^{\prime} \alpha^{\prime\prime} \alpha^{\prime\prime\prime}}\lambda_{\alpha^{\prime\prime} \alpha^{\prime\prime\prime}}   \langle \hat{B}_{\alpha^{\prime}}\rangle_{B} \Gamma^{\alpha \alpha^{\prime}}_{\alpha^{\prime\prime} \alpha^{\prime\prime\prime}}(t)
	\end{equation}
	and,
	\begin{equation}
		\chi_{\alpha\beta}(t)=t^{2}\sum_{\alpha^{\prime}\alpha^{\prime\prime} \alpha^{\prime\prime\prime}
			\beta^{\prime}\beta^{\prime\prime}\beta^{\prime\prime\prime}   }\lambda_{\alpha^{\prime\prime} \alpha^{\prime\prime\prime}} 
		\Bigl(\lambda_{\beta^{\prime\prime} \beta^{\prime\prime\prime}}\Bigl)^{\star}\langle \hat{B}_{\alpha^{\prime\prime}}\hat{B}^{\dag}_{\beta^{\prime\prime}}\rangle_{B} \Gamma^{\alpha \alpha^{\prime}}_{\alpha^{\prime\prime} \alpha^{\prime\prime\prime}}(t) \Bigl(\Gamma^{\beta \beta^{\prime}}_{\beta^{\prime\prime} \beta^{\prime\prime\prime}}(t))\Bigl)^{\star}
	\end{equation}
	Then the dissipation parameter becomes,
	\begin{equation} 
		\gamma_{\alpha\beta}=\langle \dot{\chi}_{\alpha\beta}\rangle=\frac{\chi_{\alpha\beta}(\tau)}{\tau}=\tau \sum_{\alpha^{\prime}\alpha^{\prime\prime} \alpha^{\prime\prime\prime}
			\beta^{\prime}\beta^{\prime\prime}\beta^{\prime\prime\prime}   }\lambda_{\alpha^{\prime\prime} \alpha^{\prime\prime\prime}} 
		\Bigl(\lambda_{\beta^{\prime\prime} \beta^{\prime\prime\prime}}\Bigl)^{\star}\langle \hat{B}_{\alpha^{\prime\prime}}\hat{B}^{\dag}_{\beta^{\prime\prime}}\rangle_{B} \Gamma^{\alpha \alpha^{\prime}}_{\alpha^{\prime\prime} \alpha^{\prime\prime\prime}}(\tau) \Bigl(\Gamma^{\beta \beta^{\prime}}_{\beta^{\prime\prime} \beta^{\prime\prime\prime}}(\tau))\Bigl)^{\star}
	\end{equation}
	with Lamb Shift Hamiltonian,
	\begin{equation}\label{lamb-para}
		\hat{H}_{LS}=\frac{i}{2}\sum_{\alpha}\langle \dot{\chi}_{\alpha 0}\rangle \hat{S}_{\alpha}-\langle \dot{\chi}_{\alpha 0}\rangle^{\star} \hat{S}^{\dag}_{\alpha}
	\end{equation}

\end{document}